\documentclass[twocolumn,amsmath,amssymb,aps,pra,longbibliography,floatfix]{revtex4-1}

\usepackage{physics}
\usepackage{amsmath}
\usepackage{cases}
\usepackage{url}
\usepackage{natbib}
\usepackage{textcase}
\usepackage{amssymb}
\usepackage{graphicx}
\usepackage{bm} 

\usepackage{times}
\usepackage{float}
\usepackage{multirow,microtype,color,relsize}
\usepackage{subfigure}
\usepackage[breaklinks=true]{hyperref}
\usepackage[utf8]{inputenc}
\usepackage[english]{babel}
\hypersetup{colorlinks=true,linkcolor=blue,citecolor=blue}
\hypersetup{linktocpage}
\usepackage{CJKutf8}
\usepackage{breakcites}
\usepackage{float}
\usepackage{siunitx}
\usepackage[dvipsnames]{xcolor}
\definecolor{mypine}{RGB}{1, 121, 111}

\begin{document}
\begin{CJK*}{UTF8}{gbsn}
\title{Why monovalent salt reduces charge inversion of macroion by trivalent counterions.}
\author{B. I. Shklovskii}
\affiliation{School of Physics and Astronomy, University of Minnesota, Minneapolis, Minnesota 55455, USA}
\date{\today}

\begin{abstract}
 A negative macroion, for example double helix DNA, in a 3:1 salt water solution, for example LaCl$_3$, becomes net positive due to adsorption of an excessive number of positive 3-ions. The widely accepted theory of such charge inversion is based on the idea that adsorbed at the macroion surface 3-ions form a two-dimensional strongly correlated liquid which attracts excessive 3-ions by their negative images (correlation holes). In the absence of 1:1 salt this theory reasonably agrees with experiments and numerical simulations. However, the same theory predicts that adding a large concentration of 1:1 salt increases the net positive charge, while experiments and simulations show that the charge decreases. The SCL theory aimes at the case of a very large macroion surface charge density when the effect of the competing attraction of 3-ion to its screening atmosphere in the bulk of the solution can be neglected. Here I show that for parameters used in experiments and simulations this competition makes charge inversion weak and extremely sensitive to screening of a bulk 3-ion by a large concentration of 1:1 salt. As a result charge inversion decreases with addition of 1:1 salt in agreement with experiments and simulations. This article is written for a special journal issue which celebrates important contributions to soft condensed matter physics of my friend and inspiring colleague Fyl Pincus.
\end{abstract}\maketitle
\end{CJK*}

\section{Introduction}

Charge inversion of a negatively charged macroion by a solution $Z$:1 salt is a generic phenomenon in physics, chemistry, biology and gene delivery
~\cite{Grosberg2002Colloquium,Levin2002ElectrostaticBiology,Felgner1997NonviralTherapy,Agrawal2022ElectrostaticElectrolytes,
Besteman2004DirectPhenomenon,Besteman2007ChargeIons,deserno_ovcharge,Diehl2008ColloidalConcentration,Gelbart2007DNAInspiredElectrostatics,He2009TuningInversion,
Lenz2008SimulationOvercharging,Luan_2010,lyklema2006,Nguyen2000MacroionsCharge,Perel1999ScreeningInversion,Shklovskii1999,Pianegonda2005ChargeParticles,Quesada-Perez2003OverchargingApproach,santoslevinpair2010,
Semenov2013ElectrophoreticSimulations,VanDerHeyden2006ChargeCurrents,Martin-Molina2003LookingElectrolytes,mashayakaluru2018,Naji2013Perspective:Beyond,
MoreiraStrongcouplingDistributions,Quesada-Perez2005SimulationIons,Lin2020ChargeNanopores,Motohiko2001,NGUYEN2002}.
We use this term, when a strongly negative macroion with the surface charge density $-\sigma$ attracts more positive $+Z$ ions ($Z$-ions) than necessary to neutralize it and, therefore, its surface charge density  $\sigma^*$ becomes net positive. Such a counter-intuitive phenomenon can not be explained by the Poisson-Boltzmann theory and needs a theory based on correlations between adsorbed $Z$-ions~\cite{Perel1999ScreeningInversion,Shklovskii1999,Grosberg2002Colloquium}. In this theory $Z$-ions adsorbed on the macroion surface due to their strong Coulomb repulsion form strongly correlated liquid (SCL), which reminds a Wigner crystal. When a new $Z$-ion approaches the neutral SCL it creates a negative image (correlation hole) in SCL and gets attracted to it. If concentration $N$ of $Z$:1 salt is large enough this attraction dominates the free energy loss in the bulk of solution when a new $Z$-ion joins SCL. The macroion overcharging continues until the net positive surface charge density $\sigma^*$ becomes large enough in order to stop adsorption of new $Z$-ions from the bulk. The SCL theory is also known as the strong coupling theory~\cite{MoreiraStrongcouplingDistributions}.

In the absence of additional 1:1 salt the above theory is in a reasonable agreement with experiments~\cite{VanDerHeyden2006ChargeCurrents} and numerical simulations~\cite{Lenz2008SimulationOvercharging,Quesada-Perez2005SimulationIons} for 3:1 salt. However, in the presence of a large concentration of 1:1 salt this theory~\cite{Nguyen2000MacroionsCharge} predicts that the "maximum possible" $\sigma^*$ grows with concentration of 1:1 salt $n$, while in experiments and numerical simulations $\sigma^*$ decreases with $n$.

The goal of this paper is to resolve this contradiction. The theory~\cite{Nguyen2000MacroionsCharge} 
aimed at the case of very large $\sigma$, where charge inversion is strong. Estimating the "maximum possible" charge inversion this theory
ignored the correlation free energy of screened  by monovalent salt 3-ions in the bulk of 3:1 and 1:1 salts solution. Here I adjust the theory to the same concentrations, macroion charge density $\sigma$ and ion diameters as in the Lenz and Holm (LH) simulation~\cite{Lenz2008SimulationOvercharging} including the correlation energy of screened 3-ions in the bulk.
I show that for LH parameters charge inversion by 3-ions is a marginal phenomenon very sensitive to the 1:1 salt concentration. Namely, the growth of the 1:1 salt concentration $n$ leads to weakening of charge inversion in reasonable agreement with LH simulations.

The plan of this paper is as follows. In Section II I remind and modify the capacitance model used for calculation of charge inversion of SCL in Ref.~\onlinecite{Nguyen2000MacroionsCharge}. In Section III I apply the revised theory to the set up and parameters of the Lenz and Holm (LH) simulation ~\cite{Lenz2008SimulationOvercharging} and show that apparent contradiction is now resolved.

\section{Charge inversion in mixture of 3:1 and 1:1 salt solutions}

In this section we recall results of SCL based theory~\cite{Perel1999ScreeningInversion,Shklovskii1999,Grosberg2002Colloquium} for mixture of 3:1 and 1:1 salts. We consider a large locally flat macroin with the negative charge surface density $-\sigma$. Non-linearly adsorbed on this surface 3-ions change it from $-\sigma$ to the net positive $\sigma^*$. This happens when the chemical potential of 3-ion in the bulk solution $\mu_b$ is larger than its chemical potential $\mu_s$ at the macroion surface. Below I ignore the difference in hydration energy contributions to $\mu_b$ and  $\mu_s$ as it is done in the simulation~\cite{Lenz2008SimulationOvercharging}. Then the bulk chemical potential is (we use CGS system of units)
\begin{equation}
\mu_b = -k_B T\ln(N_d/N)-(3e)^2/2Dr_s,
\label{bulk}
\end{equation}
where $T$ is the room temperature, $k_B$ is the Boltzmann constant, 
$N$ is the concentration of 3-ions, $N_d$ is the maximum geometrically possible
concentration of hydrated ions, 
\begin{equation}
r_s = (8\pi l_B n)^{-1/2}
\label{screening}
\end{equation}
is the Debye-Huckel (DH) screening radius, $l_B = e^2/Dk_BT = 0.71$ nm is 
the Bjerrum length, $D=81$ is the dielectric constant of water and $n$ is the concentration of monovalent salt. 
The second term in Eq.~\eqref{bulk} is free energy of a 3-ion interaction with its spherical DH atmosphere. 
I assume that all the ions have the same diameter $d=0.4$ nm. This leads to $N_d = d^{-3} = 26$ M (here M is used for Mol/l). 
I imagine three-dimensional lattice model with the cell $d^3$ in the bulk and two-dimensional 
lattice model with the cell $d^2$ at the macroions surface.  
 
The surface chemical potential of 3-ions $\mu_s$ is given by the sum of ideal gas and correlation terms
\begin{equation}
\mu_s = -k_B T\ln\frac{\sigma/3e}{1/d^{2}} + \mu_c =-k_B T\ln(N_d/N_s) + \mu_c,
\label{surface}
\end{equation}
where $\sigma/3e$ and $1/d^2$ are two-dimensional concentrations of 3-ions and $d^2$ cells and
$N_s=\sigma/3ed$. For used  below $\sigma = 0.95~e/$nm$^2$  and $d=0.4$ nm we get $N_s=1.4$ M.

The correlation part $\mu_c$ is given by that of the neutral one-component plasma~\cite{Perel1999ScreeningInversion,Shklovskii1999,Grosberg2002Colloquium} of 3-ions adsorbed at the negative background with charge density $-\sigma$.
It is determined by the interaction parameter 
\begin{equation}
\Gamma =\frac{(3e)^2}{DRk_BT} = \frac{9l_B}{R}, 
\label{gamma}
\end{equation}
where $R = (3e/\pi\sigma)^{1/2}$ is the radius of the disk at the macroion surface, which contains one 3-ion. For example, for used in simulation~\cite{Lenz2008SimulationOvercharging} $\sigma = 0.95~e/$nm$^2$ we get $R= 1$ nm and $\Gamma = 6.4$. Using the one component plasma fitting equation ~\cite{Totsuji1978,Shklovskii1999} 
\begin{equation}
\mu_c = -k_B T(1.65\Gamma -2.61 \Gamma^{1/4} +0.26\ln\Gamma + 1.95) 
\label{muc}
\end{equation} 
one arrives at $\mu_c =-8.8k_BT$. 

Charge inversion happens if $\mu_b > \mu_s$. In order to find resulting small positive surface charge density $\sigma^*$ I remind that $\sigma^*$ is screened by the net negative DH atmosphere at the average distance from the surface equal to the DH screening radius $r_s$.
Resulting two oppositely charged planes make a plane capacitor with the distance $r_s$ between plates and with the capacitance per unit area $C= D/4\pi r_s$. This capacitor is charged by the voltage $V=(\mu_b-\mu_s)/3e$. Then, $\sigma^* = CV$ and combining Eq.~\eqref{bulk} and Eq.~\eqref{surface} we get
\begin{equation}
\frac{\sigma^*}{\sigma} = \frac{e\ln(N/N_c)}{12\pi r_s l_B\sigma},
\label{linear}
\end{equation}
where 
\begin{equation}
N_c=N_s\exp\left(\frac{\mu_c}{k_B T} + \frac{9l_B}{2r_s}\right)
\label{critical}
\end{equation} 
is the critical concentration of 3-ions $N$ at which the macroion surface charge gets inverted.

 Eq.~\eqref{linear} was used in Ref.~\onlinecite{Nguyen2000MacroionsCharge} to estimate the maximum possible charge inversion at a very large $N \simeq N_s$ and to predict charge inversion dependence on the concentration $n$ of 1:1 salt. Ignoring the second term of the right side of Eq.~\eqref{critical} the authors of Ref.~\onlinecite{Nguyen2000MacroionsCharge} concentrated on the role of the screening radius $r_s$ in the denominator of Eq.~\eqref{linear} and claimed that at large $n$ decreasing $r_s$ leads to growth of $\sigma^*$ with growing $n$. This conclusion is correct if $\sigma$ and $|\mu_c|$ are very large, so that $N_c$ is very small and charge inversion is very strong. 

However, as I show in the next section for parameters of LH numerical simulations the second term of the right side of Eq.~\eqref{critical} cancels substantial part of $\mu_c$ and makes critical concentration $N_c$ larger and closer to $N$. This cancellation makes charge inversion weak, increases the role of screening radius in Eq.~\eqref{critical} and leads to decrease of $\sigma^*$ with growing $n$ in agreement with the LH simulations results.

\section{Comparison of theory predictions with numerical simulation results}.

In this section I compare our theory with results of numerical simulations by Lenz and Holm~\cite{Lenz2008SimulationOvercharging}. They studied a water solution of mixture of 3:1 and 1:1 salts of spherical ions with diameter $d=0.4$ nm. The large spherical macroion with the radius 5 nm, the charge $-300 e$ and the surface charge density $-\sigma = - 0.95~e/$nm$^2$ was located in the center of the cube with 22.6 nm edge. To model a solution with $N=30$ mM and no 1:1 salt the authors added 200 positive 3-ions and 300 $-e$ ones. Actually slightly more than 100 3-ions got adsorbed the macroion surface, so that remaining concentration of 3-ions in the solution was $N = 15$ mM. Below I use this $N$ for comparison with the above theory.

Lenz and Holm found that the average net charge of the macroion with adsorbed 3-ions was $+16 e$ or, in other words, they found that $\sigma^*/\sigma = 5.3\%$. 
Adding up to 1300 pairs of  $+e$ and $-e$ ion they also studied two large concentrations of 1:1 salt, $n=91$ mM and $n=196$ mM. 
They found that the average macroion excess charges were $+10 e$ and $+5e$, which correspond to $\sigma^*/\sigma=3.5\%$ and $\sigma^*/\sigma=1.7\%$ for $n=91$ mM and $n=196$ mM. These simulations clearly show that charge inversion decreases with increasing concentration of salt.

Let me compare these results with Eq.~\eqref{linear}. For two concentrations of salt $n=91$ mM and $n=196$ mM I get from Eq.~\eqref{screening} $r_s = 1$ nm and $r_s = 0.7$ nm. Then for $N_s=1.4$ M and $\mu_c =-8.8k_BT$ we get $N_c=5$ mM for $n=91$ mM and $N_c=19$ mM for $n=196$ mM. Substituting these values of $N_c$ together with $N=15$ mM into Eq.~\eqref{linear} I find that $\ln(N/N_c)= 1.1$ for $n=91$ mM and $\ln(N/N_c)= -0.24 $ for $n=196$ mM. Finally, I use Eq.~\eqref{linear} to get $\sigma^*/\sigma=4.4\%$ for $n=91$ mM and $\sigma^*/\sigma= -1.4\%$ for $n=196$ mM. 

Apparently I got a good agreement with the LH simulation result at $n=91$ mM and (marginally) lost charge inversion at $n=196$ mM. Thus, this theory agrees with observed in experiments and simulations reduction of charge inversion with increasing 1:1 salt concentration. This happens due to the important role of the second term in the exponent of Eq.~\eqref{critical}, which in turn originated from the second term of the right side of Eq.~\eqref{bulk}. It reduced the logarithm in the numerator of Eq.~\eqref{linear}. As a result, this logarithm became a faster changing function of $r_s$ and $n$ than the denominator of Eq.~\eqref{linear}. (In the theory of charge inversion of DNA by positive polyelectrolytes~\cite{NGUYEN2002} the energy of a screened by 1:1 salt polyelectrolyte molecule in the bulk of solution also played important role.) The above simple theory ignores corrections to DH screening due to finite size of ions. They can be particularly important for the largest 1:1 salt concentration $n=196$ mM.

\begin{acknowledgements}
I am grateful to A. Yu. Grosberg for useful discussions.
\end{acknowledgements}

%

\end{document}